\documentstyle[psfig,spie]{article} 

\title{The Initial Mass Function of the Most Massive Starforming Regions}

\author{Frank Eisenhauer
\skiplinehalf 
Max-Planck-Institut f\"ur extraterrestrische Physik,
Giessenbachstrasse, 85740 Garching, Germany 
}

\authorinfo{E-mail: eisenhau@mpe.mpg.de}

\begin{document} 
\maketitle 

\begin{abstract}

The stellar initial mass function (IMF) describes how many stars form
at which mass. Despite recent observational progress, many fundamental
properties of the IMF are still unknown. Specifically the question,
whether starbursts are biased towards the formation of more massive
stars, is controversially discussed in the literature. This
presentation gives an overview of how the Large Binocular Telescope
(LBT) will contribute to answering this question. I will present
(a) the status quo of the IMF research in starbursts, (b) the
importance of direct star counts in nearby templates, (c) the need for
spectroscopy, (d) the advantage of the LBT over its competitors, and
(e) what additional instrumentation I would like to see at the LBT for
the proper investigation of the most massive starforming regions.

\end{abstract}


\section{The initial mass function in massive starforming regions}

The masses of stars span at least 3 order of magnitude, from
approximately 100 solar masses for the highest mass stars to the
hydrogen burning limit of 0.1 solar masses. The relative numbers of
stars born at a given mass is described by the initial mass function
(IMF). While this distribution is of extraordinary importance for many
fields in astronomy, and while a large number of investigations have
addressed the topic, many properties of the IMF are still
controversially discussed in the literature. The fact that scientists
in the field still disagree strongly on the properties of the IMF can
be illustrated by two citations from recent reviews about the
IMF. Scalo (1998): \mbox{``... that} either the systematic uncertainties are
so large that the IMF cannot yet be estimated, or that there are real
and significant variations in the IMF index ...''; Kennicutt (1998):
``All current observations are consistent with a single universal
IMF''. Specifically, there is no consensus on the question of whether
the IMF is independent of environmental conditions such as stellar
density and metallicity. In other words: Is there a universal IMF?

Several recent investigations have come to different conclusions on
the universality of the IMF. The strongest case for the dependence of
the IMF on environmental conditions can probably be made for
starbursts. These regions of extraordinarily strong star formation
seem to be biased towards forming massive stars. Both indirect
evidence from integrated properties of extragalactic starbursts like
M82 (Rieke et al. 1993), and number counts in R 136 in 30 Doradus
(Sirianni et al. 2000) indicate that these starbursts form fewer low
mass stars than in the solar neighborhood. This hypothesis of a biased
IMF is supported by our recent investigations of NGC 3603 (Eisenhauer
et al. 1998, Brandl et al. 1999), and the HST imaging of the Arches
and Quintuplet clusters (Figer et al. 1999). Even though these
clusters are several magnitudes less massive than starburst galaxies,
and still a factor of 10 smaller than 30 Doradus, the IMF in these
clusters has a significantly shallower power law (dN/dlogM $\propto$
M$^{-0.7}$) than the IMF in the solar neighborhood (dN/dlogM $\propto$
M$^{-1.35}$, Salpeter 1955). The heating from high mass stars, and
thus the increase in the Jeans mass (Larson 1985), their radiation
fields and their strong winds may lead to this deficit in low mass
stars.  All this evidence for a bias towards massive stars in
starbursts is in contrast to the findings in the less massive galactic
OB associations (Massey et al. 1995) and young starforming regions,
which show an IMF similar to the field star population. However, the
ionizing radiation, and thus the high mass star content, of all these
optically visible clusters is by far less than what can be observed in
the most massive HII regions of our Galaxy and its closest companions.

A reliable answer to the question, whether starbursts form
preferentially higher mass stars, requires the direct detection of the
majority of the stars in these reqions. But it will be impossible in
the forseeable future to count stars with masses down to few M$_\odot$
even in the closest starburst galaxies like M82 (distance 4 Mpc). We
thus have to probe Galactic or Local Group templates.  I will define
local starburst templates as regions containing at least 100 times as
many high mass stars as Orion. This is because the IMF in such massive
clusters in the Galactic center and in NGC 3603 seems to be
significantly shallower than for example in Orion (Hillenbrand
1997). In other words, a starburst template should contain $>$ 100 O7
stars, or equivalently, its ionizing radiation should exceed 10$^{51}$
Lyman continuum photons per second. While the large distance and the
high extinction towards most giant HII regions has prevented the
detailed study of the underlying stellar population in the past, the
LBT will allow us to detect all stars in these clusters down to
sub-solar masses.

\section{From 4 m telescopes to the LBT: The galactic starburst 
template NGC 3603}

NGC 3603 is the most massive optically visible HII region of the
Galaxy, and as such has been the target of many studies. However,
detailed studies of the stellar population of NGC 3603 were hindered
by its large distance of about 7 kpc, its visual extinction of about 5
magnitudes, and its very high central concentration of about $10^5$
solar masses per cubic parsec. The central cluster of NGC 3603 can
only be resolved with state of the art telescopes and instrumentation.

\begin{figure}
\begin{center}
\begin{tabular}{c}
\psfig{figure=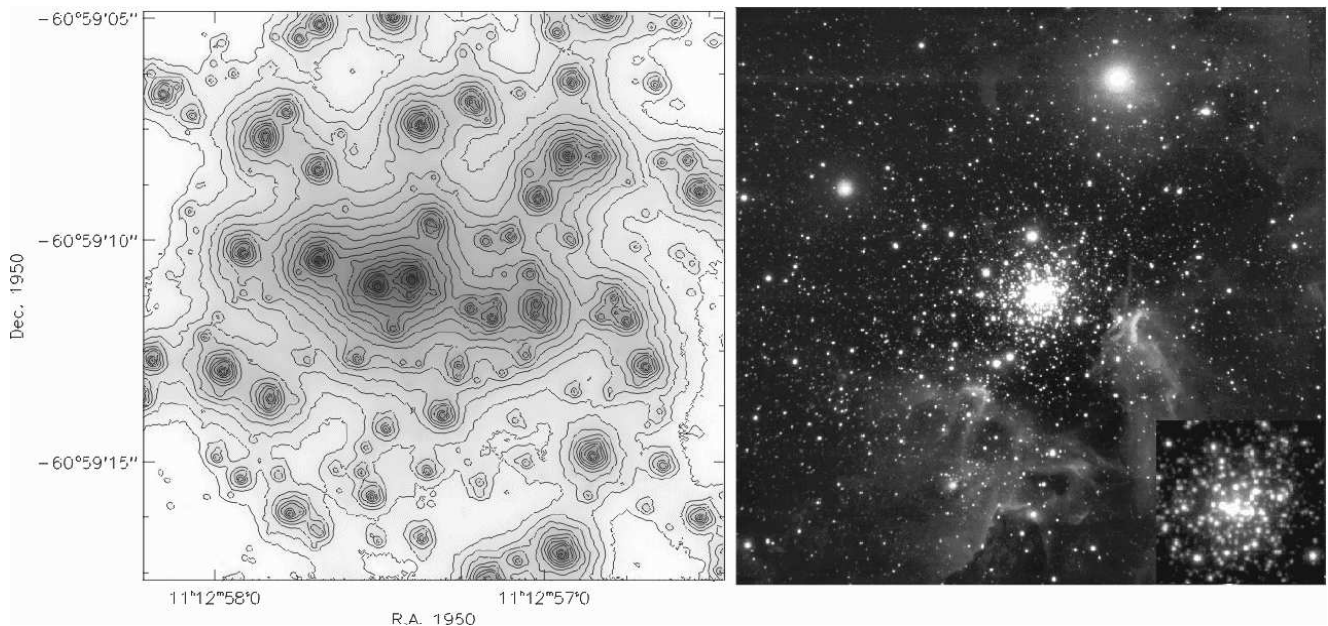,width=17.5cm}
\end{tabular}
\end{center}
\caption[NGC 3603] { \label{ngc3603} The Galactic starburst template
   NGC 3603: The left image shows a K-band map of the central regions
   of NGC 3603 obtained with the adaptive optics at the ESO 3.6 m
   telescope (Eisenhauer et al. 1998). Obviously, such high angular
   resolution images are necessary to overcome serious crowding in
   these highly compact star clusters. However, only observations with
   8 m-class telescopes are sensitive enough to detect stars with
   masses down to the hydrogen burning limit. The right image (3.4' x
   3.4') shows a near-infrared map obtained at the VLT under excellent
   seeing condition of $\approx$ 0.3'' (Brandl et al. 1999).}
\end{figure}

Optical HST imaging (Moffat et al. 1994) and ground based
speckle interferometry (Hofmann et al. 1995) have been limited
to the highest mass stars.  But progress in adaptive optics and
infrared imaging allowed us for the first time to resolve its stellar
content down to stars with masses around 1 solar mass (Eisenhauer et
al. 1998). These observations have been obtained at the ESO 3.6 m
telescope equipped with the adaptive optics ADONIS and the SHARPII+
camera (Hofmann et al. 1993). Figure \ref{ngc3603} shows the resulting
K-band image of the central cluster of NGC 3603.
While our investigations clearly excluded any turnover or truncation
in the IMF down to less than 1 solar mass, these observations with a 4
m class telescope allow no analysis of the subsolar population in NGC
3603. The faintest stars detected in J-band have been only around
19$^{th}$ magnitude. With the most recent imaging at the VLT we could
overcome this limit by about 3 magnitudes (figure \ref{ngc3603}). We
now can identify stars down to the hydrogen burning limit in NGC 3603
(Brandl et al. 1999). But even though the observations have been
carried out under excellent seeing conditions of 0.3 arcsec, severe
crowding in the central regions still prevents the statistical
analysis of these lowest mass stars. The need for higher angular
resolution images is obvious. Only adaptive optics assisted
observations at 10 m class telescopes will thus provide the answer to
the question, if the IMF in starbursts is truly deficient in low mass
stars compared to the ``universal'' field star IMF.

Such high resolution and deep infrared imaging is the unquestionable
domain of the LBT. With its twin mirrors, each 8.4 m in diameter, the
LBT has a light collecting area of almost 100 m$^2$. In addition, the
telescope will be equipped with a high order (900 elements) adaptive
optics (Salinari 2000), allowing for diffraction limited imaging at
wavelengths larger 1 $\mu$m.  Specifically important for its infrared
performance are the adaptive secondary mirrors, so that no unnecessary
thermal background is introduced by additional optical
elements. Finally, the LBT will provide a coherent beam combination of
both telescopes. This beam combiner will allow for interferometric
imaging with a maximum baseline of 23 m over a field as large as
arcminutes (Herbst 2000). The implications for deep infrared imaging
are twofold: The LBT will image 10 times fainter objects than a 4 m
class telescope, pushing the limiting magnitude in K-band below
23$^{rd}$ magnitude. The spatial resolution at this wavelength will be
better than 50 mas. We can expect to find stars with 20\%-40\% of the
mass as those seen with a 4 m class telescope (assuming \mbox{L $\propto$
M$^{1.5-2.5}$} for pre-main-sequence stars), and 3-8 times more stars
(assuming a Salpeter IMF with \mbox{N(Mass $<$ M) $\propto$
M$^{-1.35}$}). Without suffering from crowding as in our seeing limited
VLT observations of NGC 3603, we would be able to detect all stars in
NGC 3603 down to the hydrogen burning limit, and could draw a reliable
IMF in this cluster down to less than 0.1 solar masses.

\section{Hidden galactic starbursts and extragalactic HII region}

While NGC 3603 is one of the best examples to outline the scientific
driver for the determination of the IMF in starburst templates and to
highlight the advantage of the LBT over present telescopes, this
cluster (declination -61$^\circ$) can unfortunately not be observed
from Mount Graham (latitude 33$^\circ$). 

However, NGC 3603 is only one out of about a dozen clusters in the
Galaxy, which fulfill my criteria of a starburst template, having 100
times as many high mass stars as Orion. A sample of such giant HII
regions can be drawn using their total ionizing flux observed at radio
wavelengths (Georgelin \& Georgelin 1976, Smith et al. 1978).  Amongst
the most massive HII regions observable from the northern Hemisphere
are W41, W42, W43, W49 and W51. Several of these regions have been the
target of previous investigations (Blum et al. 1999, 2000, Goldader \&
Wynn-Williams 1994), but to date, the high extinction towards these
regions and the high stellar concentration have prevented the
determination of the IMF for low mass stars with 4 m telescopes.

In addition to these galactic starburst clusters, the LBT will also
allow the detailed analysis of extragalactic HII regions. While the
Magellanic Clouds are the domain of southern telescopes, the two
nearest spiral galaxies M31 and M33 are observable from Mount
Graham. Both galaxies exhibit a number of giant HII regions --- NGC
206 in M31, and NGC 595 and NGC 604 in M33 being the most prominent
examples --- and the stellar populations in these regions have been
investigated by several HST and ground based observations (Hunter \&
Winkelmann 1999, Malumuth et al. 1996, Drissen et
al. 1993). However, these observations have been limited to visible
wavelengths, and maybe a significant fraction of newly formed stars
and even the real cluster center are hidden behind several magnitudes
of extinction. Deep high resolution infrared imaging with the LBT will
thus provide more reliable measurements of the IMF in these
extragalactic HII regions.

\section{The need for spectroscopy}

Many measurements of the IMF in star clusters have been based solely
on photometry so far. By comparing multi-color information with the
theoretical stellar evolutionary tracks of post- and pre-main-sequence
stars (Meynet et al. 1994, D'Antona \& Mazzitelli 1994), one can
derive the age of the stellar population, and then convert the
measured luminosities to stellar masses. Under the best circumstances,
one can even assign an individual age and extinction to the stars. 

But specifically for the highest and lowest mass stars of young
clusters, this photometric mass determination is affected by large
uncertainties: for example, the evolution in the Hertzsprung-Russel
diagram is not single valued for stars with several tens of solar
masses. In addition, their evolutionary tracks depend highly on
individual properties like metallicity and mass loss rate, which are
not a priori known for the stellar population. Specifically, the mass
loss rate can vary strongly from star to star. The situation for low
mass stars in their early pre-main-sequence evolution is no
better. Still partially hidden in their dust cocoons, the extinction
can vary dramatically from one star to the next. These stars also
exhibit strong excess emission from surrounding accretion disks, which
adds significantly to the total flux in the infrared. We therefore
require spectroscopy for the proper mass determination of the high and
low mass stars in the young starforming regions.  But optical
spectroscopy for the stars in many starburst regions is prevented by
the high extinction, and we have to restrict ourselves to infrared
wavelengths.

Near infrared spectroscopy can indeed provide the diagnostic
tools. Given sufficient spectral resolution of 2000 and a signal to
noise ratio of 30, derived stellar parameters such as mass-loss rate,
luminosity, surface abundance and temperature show good agreement
between optical and near infrared spectral analysis of high mass Of
and WNL stars (Bohannan \& Crowther 1999). The effective temperature
of low-mass-stars can be determined to approximately \mbox{$\pm 300$
K} from several spectral features (e.g. CaI, NaI, $^{12}$CO(2,0)) in
the K-band, which vary significantly with temperature (Ali et
al. 1995). The spectral resolution in the analysis of Ali et
al. (1995) was approximately 1400.

This need for advanced spectroscopic capabilities brings me to my
wish-list for additional instrumentation for the LBT.

\section{Additional instrumentation for the LBT}

An accurate measurement of the initial mass function from direct star
counts requires infrared imaging and spectroscopy of several hundred
to thousand stars at the highest angular resolution. The LBT with its
high order adaptive optics provides the basis for such investigations.

However, almost all giant galactic starforming regions are hidden
behind so many magnitudes of extinction that no sufficiently bright
reference star is within the isoplanatic patch to correct for
atmospheric aberrations using adaptive optics. There are two possible
techniques to overcome this problem. The first possibility is the
installation of a laser guide star facility, which effectively will
provide full sky coverage. Such a facility was integrated and tested
for example at the Calar Alto Observatory (Eckart et al. 2000). The
second possibility would be the implementation of an infrared wave
front sensor. While such a wave front sensor is only of minor help for
extragalactic astronomy, the hidden galactic clusters have
sufficiently bright stars to be used as a wave front reference in the
infrared. Such an infrared wave front sensor is presently implemented
in the adaptive optics at the VLT (Rousset et al. 1998).

In terms of its spectroscopic capabilities, the first generation of
instruments at the LBT is not well suited for our purpose. The near
infrared camera and spectrometer LUCIFER (Mandel 2000) will provide
multi-object spectroscopy only for seeing limited observations. The
use of its integral field unit or of a long slit will not allow the
spectroscopy of several hundred to thousand stars in crowded
fields. Such a detailed analysis of the stellar population in
starbursts will require a cryogenic multi-object spectrograph working
at the diffraction limit of the LBT.

In summary, I would like to see a laser guide star
facility or an infrared wave front sensor, and a diffraction limited
multi-object spectrograph at the LBT as soon as possible.

\section{Summary}

The most massive star forming regions in our Galaxy and in the nearby
spirals M31 and M33 are prime targets for a direct measurement of
the IMF. These observations will provide a reliable test of hypothesis
that the IMF is biased towards massive stars in starburst, and will
thus give convincing evidence for or against a universal IMF.
Optimized for infrared observations, providing a large collecting
area, and equipped with a high order adaptive optics, the LBT will
contribute significantly to the measurement of the IMF in these dense
and distant clusters. However, a laser guide star facility or an
infrared wave front sensor, and a diffraction limited multi-object
spectrograph should be installed at the LBT, to allow for the observation
of hidden clusters and more reliable mass estimates for the individual
stars.

\section*{Bibliography}

Ali, Carr, Depoy, Frogel \& Sellgren 1995, AJ, 110, 2415 \\
Bohannan \& Crowther 1999, ApJ, 511, 374 \\
Blum, Damineli \& Conti 1999, AJ, 117, 1392 \\
Blum, Conti \& Damineli 2000, AJ, 119, 1860 \\
Brandl, Brandner, Eisenhauer, Moffat, Palla \& Zinnecker 1999, A\&A,
352, L69  \\ 
D'Antona \& Mazzitelli 1994, ApJS, 90, 467 \\
Drissen, Moffat \& Shara 1993, AJ, 105, 1400 \\
Eckart, Hippler, Glindemann et al. 2000, ExA, 10, 1 \\
Eisenhauer, Quirrenbach, Zinnecker \& Genzel 1998, AJ, 498, 278 \\
Figer, Kim, Morris, Serabyn, Rich \& McLean 1999, ApJ, 525, 750  \\
Georgelin \& Georgelin 1976, A\&A, 49, 57 \\
Goldader \& Wynn-Williams 1994, ApJ, 433, 164 \\
Herbst 2000, this conference \\
Hillenbrand 1997, 113, 1733 \\
Hofmann, Seggewiss \& Weigelt 1995, A\&A, 300, 403 \\
Hofmann, Brandl, Eckart, Eisenhauer, Tacconi-Garman 1995, SPIE,
2475, 192 \\ 
Hunter \& Winkelman 1990, PASP, 102, 854 \\
Larson 1985, MNRAS, 214, 379 \\
Malumuth, Waller \& Parker 1996, AJ, 111, 1128 \\
Mandel 2000, this conference \\
Massey, Johnson \& Degioia-Eastwood 1995, ApJ, 454, 151 \\
Moffat, Drissen \& Shara 1994, ApJ, 636, 183 \\
Meynet, Maeder, Schaller, Schaerer \& Charbonnel 1994, A\&AS, 103, 97 \\
Kennicutt 1998, ASP Conference Series, 142, 1 \\
Rieke, Loken, Rieke \& Tamblyn 1993, ApJ, 412, 99 \\
Rousset, Lacombe, Puget et al. 1998, SPIE, 3353, 508 \\
Salinari 2000, this conference \\
Salpeter 1955, ApJ, 121, 161 \\
Scalo 1998, ASP Conference Series, 142, 201 \\
Sirianni, Nota, Leitherer, De Marchi \& Clampin 2000, ApJ, 533, 203 \\
Smith, Mezger \& Biermann 1978, A\&A, 66, 65 \\

\end{document}